\documentclass[aps,prl,twocolumn,superscriptaddress,showpacs,amsmath,amssymb, hidelinks]{revtex4-1} 
\usepackage{comment}
\usepackage{orcidlink}
\usepackage{graphicx} 
\usepackage{epstopdf} 
\usepackage{dcolumn} 
\usepackage{xcolor} 
\usepackage{amssymb} 
\usepackage{hyperref} 
\usepackage[utf8]{inputenc} 
\usepackage[T1]{fontenc}
\usepackage[english]{babel} 
\usepackage[displaymath, mathlines]{lineno} 
\usepackage{blindtext} 
\usepackage{amsmath}
\usepackage{xcolor}
\usepackage{siunitx}
\usepackage{relsize}

\graphicspath{{figures/}} 

\newcommand {\mugammarec} {$e^{+}e^{-}\rightarrow \mu^+ \mu^- \gamma$}

\newcommand {\qq} {$e^{+}e^{-}\rightarrow q \bar q (\gamma)$}

\newcommand {\ztautau} {$e^{+}e^{-} \rightarrow Z^\prime \, (\rightarrow \tau^+\tau^-) \ \mu^+ \mu^-$}
\newcommand {\xtautau} {$e^{+}e^{-}\rightarrow X \, ( \rightarrow \tau^+\tau^-) \ \mu^+ \mu^-$}
\newcommand {\xtautauBF} {$\sigma(e^{+}e^{-}\rightarrow \mu^+ \mu^- X)\times \mathcal{B}(X \rightarrow \tau^+\tau^-)$}
\newcommand{\sigmaXtautau}{$\sigma(e^+e^- \to \ X \ (\to \tau^+ \tau^-) \ \mu^+\mu^-)$}

\newcommand {\tgamma} {$e^{+}e^{-}\rightarrow \tau^+ \tau^- (\gamma)$}
\newcommand {\eemumu} {$e^{+}e^{-}\rightarrow e^{+}e^{-} \mu^+\mu^-$}
\newcommand {\fourmu} {$e^{+}e^{-}\rightarrow \mu^{+}\mu^{-} \mu^+\mu^-$}
\newcommand {\eett} {$e^{+}e^{-}\rightarrow e^{+}e^{-} \tau^+\tau^-$}
\newcommand {\eeee} {$e^{+}e^{-}\rightarrow e^{+}e^{-} e^+ e^-$}
\newcommand {\mmtt} {$e^{+}e^{-}\rightarrow \mu^{+}\mu^{-} \tau^+\tau^-$}

\newcommand {\eepp} {$e^{+}e^{-}\rightarrow e^{+}e^{-} \pi^+\pi^-$}
\newcommand {\eex} {$e^{+}e^{-}\rightarrow e^{+}e^{-} h$}
\newcommand {\mumupp} {$e^{+}e^{-}\rightarrow \mu^{+}\mu^{-} \pi^+\pi^-$}

\newcommand {\tautau} {$\tau^+\tau^-$}

\newcommand {\lmultau} {$L_{\mu}-L_{\tau}$}

\newcommand {\zprime} {$Z^\prime$}

\newcommand {\gevcc}{GeV/$c^2$}
\newcommand {\mevcc}{MeV/$c^2$}

\newcommand {\mrec}{$M_{\text{recoil}}(\mu\mu)$}
\newcommand {\mtrk}{$M({\text{4~tracks}})$}

\def\babar{\mbox{\slshape B\kern-0.1em{\smaller A}\kern-0.1em B\kern-0.1em{\smaller A\kern-0.2em R}}}

\begin{document}

{\onecolumngrid
\noindent
KEK preprint: 2022-32\\
Belle II preprint: 2022-006\\
}

\title{\protect Search for a \texorpdfstring{$\tau^+\tau^-$}{tau} resonance in \texorpdfstring{\mmtt}{mmtt} events  with the Belle~II experiment}

  \author{I.~Adachi\,\orcidlink{0000-0003-2287-0173}} 
  \author{K.~Adamczyk\,\orcidlink{0000-0001-6208-0876}} 
  \author{L.~Aggarwal\,\orcidlink{0000-0002-0909-7537}} 
  \author{H.~Ahmed\,\orcidlink{0000-0003-3976-7498}} 
  \author{H.~Aihara\,\orcidlink{0000-0002-1907-5964}} 
  \author{N.~Akopov\,\orcidlink{0000-0002-4425-2096}} 
  \author{A.~Aloisio\,\orcidlink{0000-0002-3883-6693}} 
  \author{N.~Anh~Ky\,\orcidlink{0000-0003-0471-197X}} 
  \author{D.~M.~Asner\,\orcidlink{0000-0002-1586-5790}} 
  \author{H.~Atmacan\,\orcidlink{0000-0003-2435-501X}} 
% ...
\collaboration{The Belle II Collaboration}

\begin{abstract}

We report the first search for a non-standard-model resonance decaying into $\tau$ pairs 
in \mmtt\ events in the 3.6--10~\gevcc\ mass range. 
We use a 62.8~fb$^{-1}$ sample of $e^+e^-$ collisions collected at a center-of-mass energy of 10.58~GeV by the Belle~II experiment at the SuperKEKB collider. 
The analysis probes three different models predicting a spin-1 particle coupling only to the heavier lepton families, a Higgs-like spin-0 particle that couples preferentially to charged leptons (leptophilic scalar), and an axion-like particle, respectively.
We observe no evidence for a signal and set exclusion limits at 90\% confidence level on the product of cross section and branching fraction 
into $\tau$ pairs, ranging from 0.7~fb to 24~fb, and on the couplings of these processes. We obtain world-leading constraints on the couplings for the leptophilic scalar model for masses above 6.5 GeV/$c^2$ and for the axion-like particle model over the entire mass range.

\end{abstract}

\maketitle

Despite its successes, the standard model (SM) of particle physics is known to provide an incomplete description of nature. 
For example, it does not  address the phenomenology related to the existence of dark matter~\cite{BERTONE2005279}, specifically in the prediction of the  observed relic density. In addition, experimental observations showed inconsistencies with the SM. 
Prominent examples are the long-standing difference between the measured and the expected value of the muon anomalous magnetic-moment~\cite{PhysRevD.73.072003, PhysRevLett.126.141801, AOYAMA20201}, and the tensions in flavor observables reported by the \babar, Belle, and LHCb experiments~\cite{BaBar:2013mob, LHCb:2017rln, Belle:2019rba, PhysRevD.108.032002, PhysRevD.106.L031703, SALA2017205, Chen:2017usq, Greljo:2021xmg}.
Some of these observations can be explained with the introduction of additional, possibly lepton-universality-violating interactions mediated by non-SM neutral bosons. 
Examples are the  \lmultau\ extension of the SM, a Higgs-like spin-0 particle (leptophilic scalar),  
and axion-like particles (ALPs).
The \lmultau\ model
gauges the difference between the muon and the $\tau$-lepton numbers through the introduction of a neutral spin-1 
boson \zprime\ that couples only to the second and third generations of leptons~\cite{PhysRevD.43.R22,Shuve:2014doa, Altmannshofer:2016jzy}. The \zprime\ could also mediate interactions between SM and dark matter. The leptophilic scalar $S$ 
is an hypothetical particle that couples preferentially to charged leptons through a parameter $\xi$ and Yukawa-like couplings to the individual families proportional to the lepton masses~\cite{batell2017muon}.  
Axion-like particles appear in many models with spontaneous breaking of global symmetries as relics of high-energy extensions of the SM~\cite{Bauer_2017,bauer2022flavor}.  
In some models, they couple to charged leptons through parameters $C_{\ell\ell}$ with $\ell = e, \mu, \tau$, 
with a decay rate 
to leptons proportional to the squared lepton-masses. The coupling to charged leptons is parametrized as $|C_{\ell\ell}|/\Lambda$, 
where $\Lambda$ is the scale of the global symmetry breaking~\cite{bauer2022flavor}.
For the ALP model we follow the approach of Refs.~\cite{Bauer_2017,bauer2022flavor}, in which the coupling of ALPs to charged leptons is studied assuming no coupling to all the other particles, in particular photons.

Searches for a $Z^{\prime}$ decaying to muons have been reported by the \babar, Belle, and CMS collaborations~\cite{TheBABAR:2016rlg, PhysRevD.106.012003, cms}. 
An invisibly decaying \zprime\ has been searched for by  the Belle~II~\cite{PhysRevLett.124.141801,PhysRevLett.130.231801}
and NA64-$e$~\cite{PhysRevD.106.032015} experiments. 
The leptophilic scalar decaying into electrons and muons is constrained by \babar\  for masses up to approximately 6.5~\gevcc~\cite{PhysRevLett.125.181801}.
Decays of ALPs into leptons are constrained mostly through reinterpretations of other measurements~\cite{Bauer_2017,bauer2022flavor}.
For all these particles, decays into pairs of $\tau$ leptons are unconstrained due to the experimental difficulties in fully reconstructing the final state that has multiple neutrinos.

In this Letter, we search for a $X \to \tau^+ \tau^-$ resonance, where $X = Z^\prime, S$, or ${\rm ALP}$, in \mmtt\ events. The signal signature is a narrow enhancement in the distribution of the recoil mass against two oppositely charged muons \mrec. This exploits the fact that in electron-positron colliders the beam energy is entirely transferred to the final-state collision products.
We use a sample of $e^{+}e^{-}$ collisions produced at a center-of-mass (c.m.) energy $\sqrt{s}=10.58$~GeV in 2019--2020 by the SuperKEKB asymmetric-energy collider~\cite{superkekb} at KEK. The data, recorded by the Belle II detector~\cite{Abe:2010sj, ref:b2tip}, correspond to an integrated luminosity of 62.8~fb$^{-1}$~\cite{lumi}.
The \lmultau\ model is used as a benchmark to devise the analysis selections through the process \ztautau. 
We then check the selection performance on the two additional models. In all the cases,  the $X$ resonance  is predominantly emitted as final-state radiation (FSR) off one of the two muons. We restrict our analysis to $\tau$-lepton decays 
to one charged particle  
and any number of neutral particles. 
We therefore select events with exactly four charged particles, where at least two are identified as muons (tagging muons). 
The main expected backgrounds are the processes \qq\ 
with $q = u,d,s,c,b$, \tgamma\ 
where one $\tau$ lepton decays into a one-charged-particle state (one-prong) 
and the other to a three-charged-particle state (three-prong), and four-lepton processes \eemumu, \mmtt, and \eett.
A multivariate analysis exploits kinematic variables to enhance the signal-to-background ratio. A control sample is used to compare data to simulation, from which the most important systematic uncertainties are estimated.
The signal yield is extracted 
through fits to the \mrec\ distribution, which allows an estimate of the background directly from data. 
To minimize bias, the analysis techniques are defined using simulated events prior to examining data.

The Belle~II detector~\cite{Abe:2010sj, ref:b2tip} consists of several subdetectors arranged in a cylindrical structure around the $e^{+}e^{-}$ interaction point. 
The longitudinal direction, the transverse plane, and the polar angle $\theta$ are defined with respect to the detector's cylindrical axis in the direction of the electron beam. Charged-particle trajectories (tracks) are reconstructed by a tracking system consisting of a two-layer silicon-pixel detector, surrounded by a four-layer double-sided silicon-strip detector and then a central drift chamber (CDC) covering
$17^{\circ}$$<$$\theta$$<$$150^{\circ}$. 
The second pixel layer was only partially installed for the data sample we analyze, covering one sixth of the azimuthal angle. Outside the CDC, time-of-propagation and aerogel ring-imaging Cherenkov detectors cover $31^{\circ}$$<$$\theta$$<$$128^{\circ}$ and $14^{\circ}$$<$$\theta$$<$$30^{\circ}$, 
respectively, to provide charged-particle identification. The electromagnetic calorimeter (ECL) reconstructs photons and identifies electrons in the range $12^{\circ}$$<$$\theta$$<$$155^{\circ}$.
It fills the remaining volume inside a superconducting solenoid that generates a 1.5-T field.  
A $K^{0}_{L}$ and muon detection subsystem (KLM) is installed in the iron flux return of the solenoid and covers  
$18^{\circ}$$<$$\theta$$<$$155^{\circ}$.

The identification of muons relies mostly on charged-particle penetration in the KLM for momenta larger than 0.7~GeV/$c$ and on information from the CDC and ECL otherwise. 
Electrons are identified mostly by comparing measured  momenta  with energies of the associated ECL depositions. We identify charged hadrons 
as particles not compatible with both electrons and muons. 
Charged pions are identified combining the information from all subdetectors except the silicon detectors. Photons are reconstructed from ECL-energy depositions 
greater than 100~MeV not associated with any track. Neutral pions are identified as pairs of photons with an invariant mass within three standard deviations from the known $\pi^0$ mass. Details of particle reconstruction and identification are given in Refs.~\cite{ref:b2tip, tracking}. 

Signal events are simulated using \texttt{\textsc{MadGraph5@NLO}}~\cite{Alwall2014}, including initial-state radiation (ISR). 
The \mrec\ resolution varies with the \zprime\ mass: it is 30~\mevcc\ at the kinematic threshold 2$m_{\tau}$ and decreases smoothly to 10~\mevcc\ at 6~\gevcc\ and to 1~\mevcc\ at 10~\gevcc. We generate events for $Z^{\prime}$ masses ranging from 3.6 to 10~\gevcc\  in steps of 25, 20, 10, and 5~\mevcc, following the \mrec\ resolution. The background processes are simulated using the following generators:
$e^{+}e^{-} \rightarrow u \bar u, d \bar d, s \bar s, c \bar c$ with \texttt{KKMC}~\cite{jadach2000precision} interfaced with \texttt{\textsc{Pythia8}}~\cite{pythia8} and \texttt{\textsc{EvtGen}}~\cite{evtgen};
$e^{+}e^{-} \rightarrow b\bar b$\ with \texttt{\textsc{EvtGen}}; \tgamma\ with \texttt{KKMC} interfaced with \texttt{\textsc{Tauola}}~\cite{ref:tauola};
\eemumu, 
\mmtt,
\eett,
\fourmu,
and \eeee\ with \texttt{AAFH}~\cite{ref:fourlepton};
\eepp\ with \texttt{\textsc{Treps}}~\cite{uehara2013treps}; and
$e^{+}e^{-}\rightarrow \mu^+ \mu^- (\gamma)$\ with \texttt{KKMC}. Electromagnetic FSR is simulated with {\texttt{\textsc{Photos}}}~\cite{Barberio:1990ms, Barberio:1993qi} for processes generated with {\tt \textsc{EvtGen}}. All the four-lepton processes generated with \texttt{AAFH} do not include ISR effects.
Additional non-simulated backgrounds include \mumupp\ processes and two-photon processes \eex, where $h$ is typically a low-mass hadronic system. The detector geometry and interactions of final-state particles with detector material are simulated using \texttt{\textsc{Geant4}}~\cite{ref:geant4} and the Belle~II software~\cite{basf2, basf2-zenodo}. 

The online event selection (trigger) is a logical OR of a three-track trigger and a single-muon trigger. The former 
requires the presence of at least three tracks in $37^\circ$$<$$\theta$$<$$120^\circ$. 
The latter is based on the match between CDC tracks and signals in the 
$51^\circ$$<$$\theta$$<$$117^\circ$ KLM polar range. An unbiased measurement of the efficiency of both triggers is performed 
using a reference trigger, which requires that the total ECL-energy deposition in $22^{\circ}$$<$$\theta$$<$$128^{\circ}$
exceeds 1~GeV. This is achieved by requiring the presence of one electron with energy above 1~GeV.
The three-track-trigger efficiency is 
measured in four-track events containing at least two pions and one electron. The single-muon-trigger efficiency is measured in events with one electron and one muon: the efficiency for events with multiple muons is computed using the single-muon efficiency, assuming no correlation. 
The overall trigger efficiency is 96\% for \zprime\ masses up to 8~\gevcc, then it decreases smoothly to 90\% at  9~\gevcc\ and to 50\% at 10~\gevcc.

To suppress misreconstructed and beam-induced background tracks, we require  that the transverse and longitudinal projections of their distance of closest approach to the  interaction point be smaller than 0.5 and 2.0~cm, respectively.
We require that events have exactly four charged  particles with zero net charge, with at least a pair  of oppositely-charged particles identified as muons and the remaining two particles separately identified as electrons, muons, or charged hadrons, for which we assume the electron, muon, and pion mass hypotheses, respectively. 
Events with more than two muons produce up to four candidates. 
We require that the  four-track invariant mass \mtrk\ be below 9.5~\gevcc\ to suppress the four-lepton backgrounds that peak at the c.m. energy, such as \eemumu, \fourmu, and \eeee. The remaining background is largely dominated by \qq\ and \tgamma\ and, to a lesser extent, by \eemumu\ processes.  
The final signal-from-background discrimination relies on signal-event properties: 
presence of a resonance recoiling against the two tagging muons;  
FSR emission of the resonance;
and compatibility of the system recoiling against the tagging muons with a \tautau\ pair. We identify 14 variables~\cite{supplemental_reference}, among which the most discriminating are the following: the momenta of the two tagging muons in the c.m.~frame; 
the components of the recoil momentum (the \zprime\ momentum, for signal events) transverse to the momentum direction of each of the two tagging muons in the c.m.~frame; 
and topological variables defined in the rest frame of the system recoiling against the two muons, such as the thrust value \cite{PhysRevLett.39.1587, Brandt:1964sa}, the first Fox-Wolfram moment shape-variable~\cite{fox-wolfram}, and the angles between the thrust direction and the directions of the muons. 
We preprocess some variables to reduce their dependence on the $Z'$ mass.
For example, the momentum variables are scaled by the maximum  momentum of the system with mass \mrec\  
recoiling against the two tagging muons. 

We use multilayer perceptrons (MLPs)~\cite{hoecker2009tmva}, trained on simulated signal and background events, 
with 14 input nodes and one  output node 
for the signal-from-background discrimination. 
To improve performance, we use eight separate MLPs in different \mrec\ intervals, which we refer to as MLP ranges, approximately 1~\gevcc\ wide. 
The selection applied on the node
output is optimized separately in each MLP range  with a figure of merit~\cite{Punzi:2003bu} and then 
expressed as a function of \mrec\ by interpolation. 
The resulting signal efficiency varies with the \zprime\ mass from 12\% near the kinematic threshold  $2m_\tau$ to 2\% at 10~\gevcc~\cite{supplemental_reference}. 
The background suppression reduces the \qq\ and \tgamma\ processes by two to three orders of magnitude, so that the resulting expected background contains significant contributions from the four-lepton \mmtt\ and \eett\ processes, which were small before the MLP selection. The fraction of surviving events with more than one candidate is negligible. 

We apply the full selection on signal events simulated according to the two additional models, and compare the signal efficiencies with those estimated for the \lmultau\ model:  
relative differences are in the range 10\%--20\%. 

\begin{figure}[!htb] 
  \centering
    \includegraphics[width=0.98\linewidth]{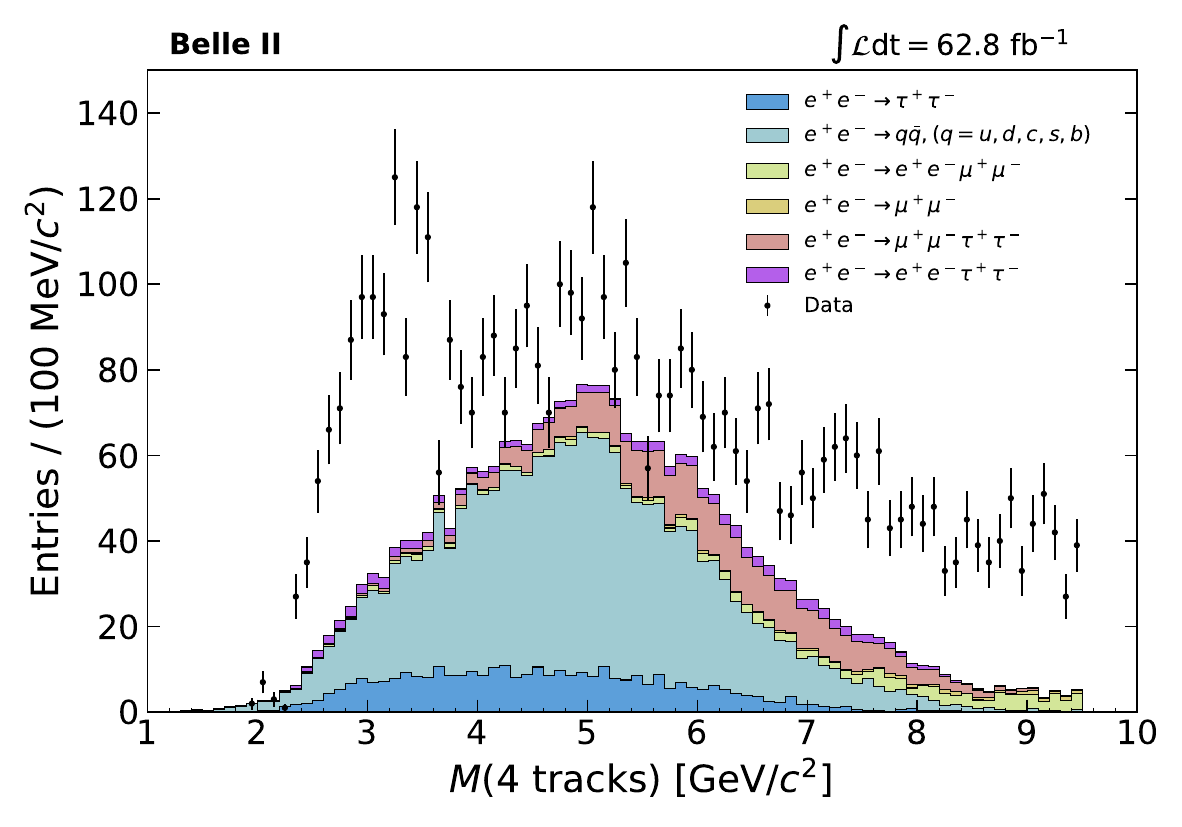}
    \caption{Observed four-track invariant mass distribution compared to the expectations of the simulation. Contributions from the various simulated processes are stacked.} 
    \label{fig:4track}
\end{figure}

The \mtrk\ distribution after all selections applied is compared with the simulation in Fig.~\ref{fig:4track}.
The discrepancies between data and simulation are due to  large contributions from non-simulated two-photon processes \eex\  for  \mtrk~$<$~4~\gevcc, and to missing ISR in simulated four-lepton processes for \mtrk~$>7$~\gevcc.
Additional contributions to the observed discrepancies come from the process \mumupp. The origin of these discrepancies is confirmed by specific studies on the \mtrk\ distribution before the MLP selection~\cite{supplemental_reference} and on a  control sample after all the selections. 
A pion-tagged control sample is selected by applying the analysis requirements with the two tagging muons replaced by two charged pions.  
Both samples are dominated by  \qq\ and \tgamma\ processes, which include ISR in the respective generators. 
In both cases we observe good agreement for dimuon or dipion masses greater than 2~\gevcc, where the two-photon processes \eex\ do not contribute~\cite{supplemental_reference}. 

\begin{figure}[!htb] 
  \centering
    \includegraphics[width=0.98\linewidth]{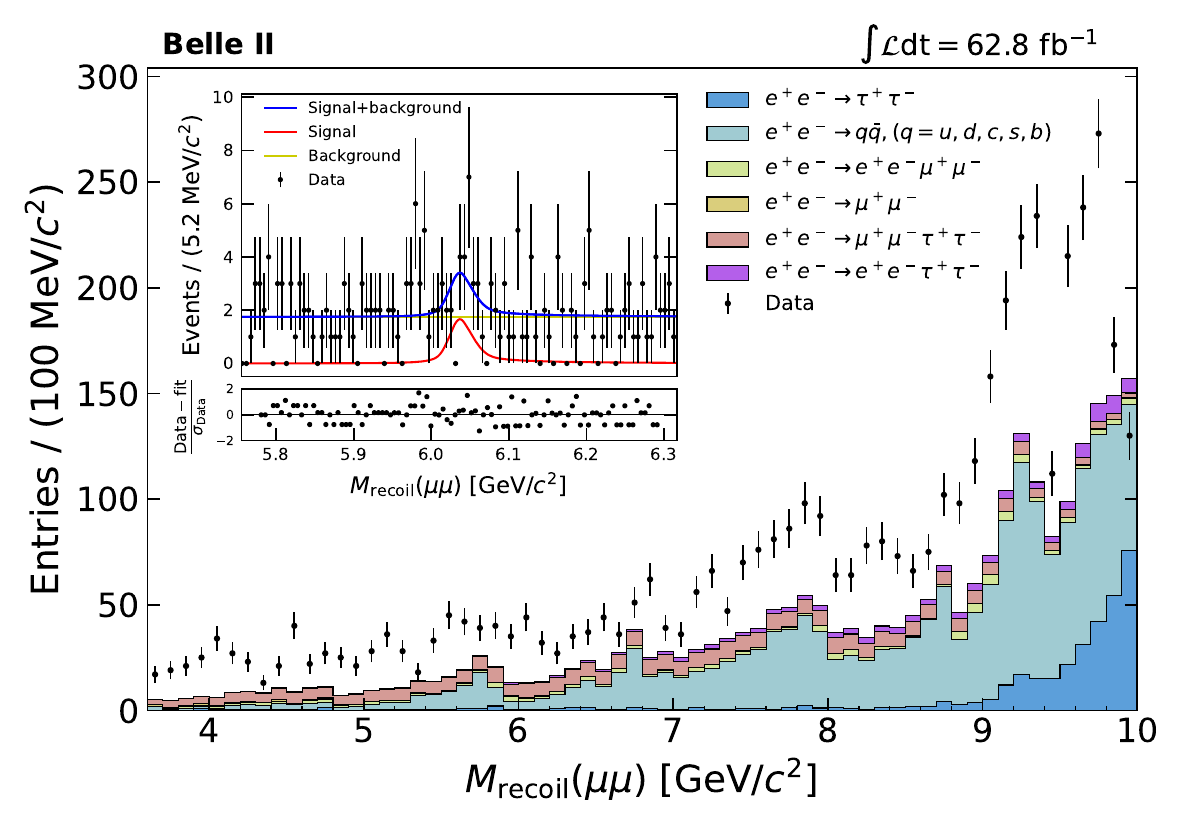}
    \caption{Observed distribution of the recoil mass against the two tagging muons, compared to the expectations of the simulation. Contributions from the various simulated processes are stacked. The inset shows an example fit at a signal mass hypothesis of 6.036~\gevcc, and the difference between the number of observed and fitted events, 
    divided by the statistical uncertainty of the former.} 
    \label{fig:mrecoil}
\end{figure}

The \mrec\ distribution after all the selections are applied is shown in Fig.~\ref{fig:mrecoil}. Discrepancies induced by the lack of ISR effects in four-lepton simulation appear mainly for \mrec\ below approximately 6~\gevcc.
Above 9~\gevcc\ the discrepancies are due to two-photon \eex\ processes. Also visible are variations among the eight MLP ranges. 
Neither of these effects produce narrow peaking structures at the scale of the signal resolution 
in the \mrec\ distribution, as shown by the inset in  Fig.~\ref{fig:mrecoil}.  

The signal yields are obtained from a scan over the \mrec\ spectrum through a series of unbinned maximum likelihood fits.
The signal \mrec\ distributions are parametrized from the simulation as sums of two Crystal Ball functions~\cite{Skwarnicki:1986xj} sharing the same mean value. 
The  scan step-size is half the mass resolution. Each fit extends over an interval 40 times larger than the \zprime\ mass resolution. 
The background is described with a constant. Higher-order polynomials for the background parametrization are investigated, but their  coefficients are compatible with zero over the full recoil-mass spectrum. A total of 2384 fits are performed, covering the range 3.6--10~\gevcc. If a fitting interval extends over two different MLP ranges, we use data selected by the MLP corresponding to the range where the central mass value is located.
The fit determines the signal and background yields using a fixed signal shape.
We then convert signal yields into cross sections, after correcting for signal efficiency and luminosity. 

Several sources of systematic uncertainties affecting the cross-section determination are taken into account: they are related to signal efficiency, luminosity, and fit procedure. 
Uncertainties due to the trigger efficiency are evaluated by propagating the uncertainties on the measured trigger efficiencies. The relative uncertainty on the signal efficiency is  2.7\% across the entire mass range. 
Uncertainties due to the tracking efficiency are estimated in \tgamma\ events, in the one-prong against three-prong topology. The relative uncertainty on the signal efficiency is 3.6\%.
Uncertainties due to the particle-identification requirement are studied using \mugammarec, \eemumu, \eeee, \eepp events and final states with either a $J/\psi$ or a $K^0_S$. 
The relative uncertainty on the signal efficiency varies between 3.9\% and 6.2\%, depending on the \zprime\ mass.
Uncertainties due to the MLP selection efficiency are evaluated on 
the pion-tagged control sample. 
We compare MLP  
efficiencies in data and simulation in signal-like regions of the  control sample and assume that uncertainties estimated in those conditions are representative of the signal conditions.
We find good agreement between data and simulation and estimate a 2.8\% relative uncertainty on the signal efficiency from the uncertainty of the data-simulation comparison.
Uncertainties due to the interpolation of the signal efficiency  between simulated mass points are  2.5\%, which is assigned as a relative uncertainty on the signal efficiency.
Uncertainties due to the fit procedure are evaluated using a bootstrap technique~\cite{efron1994introduction}.
Signal events from simulation are overlaid on simulated background with a yield corresponding to the excluded 90\% CL value and fitted for each \zprime\ mass. 
The distribution of the difference between the overlaid and the fitted yields, divided by the fit uncertainty, has a negligible average bias with a width that deviates from one by 4\%, which is assigned as a relative uncertainty on the signal-yield determination.
Uncertainties due to differences in the recoil-mass resolution between data and simulation are evaluated by introducing an additional smearing on the simulated momenta of the two tagging muons, which reflects the difference in momentum resolution measured with cosmic rays and in  $D^{\ast +} \to D^0 \pi^+$ decays with respect to the simulation predictions. The relative uncertainty on the signal-yield determination is 3\%. 
The relative uncertainty on the signal efficiency due to the knowledge of the beam energy is 1\%~\cite{PhysRevD.108.032006}. The uncertainty due to the selection on the four-track invariant mass is negligible. 
Finally, a relative uncertainty of 1\% on the integrated luminosity is considered~\cite{lumi}.

All the systematic uncertainties are summed in quadrature: the final relative systematic uncertainty on the cross section varies in the range 8.8\%--10.0\% depending on the \zprime\ mass. We account for systematic uncertainties by approximating their effects as a Gaussian smearing of the signal efficiency.

\begin{figure}[!htb] 
  \centering
    \includegraphics[width=0.98\linewidth]{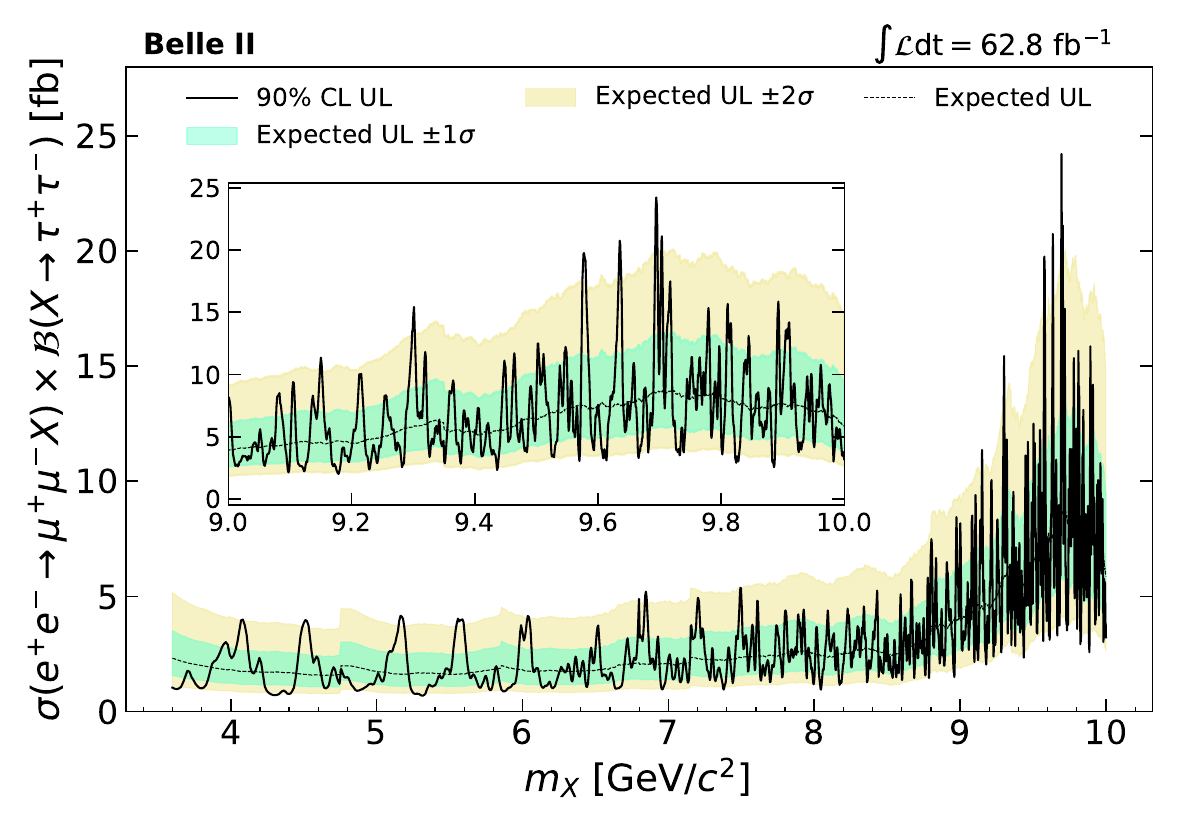}
    \caption{Observed 90\% CL upper limits and corresponding expected limits on the cross section for the process \xtautau\ 
    with $X = Z^\prime, S, {\rm ALP}$
    as functions of the $X$ resonance mass. The inset shows a magnification of the region above 9~\gevcc.} 
    \label{fig:xs}
\end{figure}

The significance is evaluated as $ \sqrt{2\, \text{log} (\mathcal{L}/\mathcal{L}_{0})}$
where $\mathcal{L}$ and $\mathcal{L}_{0}$ are the likelihoods of the fits with and without signal. 
The largest local significance observed is $3.0 \, \sigma$, corresponding to a global significance of $1.8 \, \sigma$, at a recoil mass of 9.695~\gevcc~\cite{supplemental_reference}.  
Since we do not observe any significant excess above the background, we derive 90\% CL upper limits on the process cross section \sigmaXtautau\ $=$ \xtautauBF\ with $X = Z^\prime, S, {\rm ALP}$, 
using the frequentist procedure $\rm CL_S$~\cite{Cowan:2010js, CLS}. The limits are shown in Fig.~\ref{fig:xs}. 
Expected limits are defined as median limits from background-only simulated samples that use background yields observed from the fits to data.  
The combination of the variations originating from the MLP ranges and of the overlap between the fit intervals induces an oscillatory behaviour. 
The resulting upper limits are dominated by sample size, with systematic uncertainties worsening them on average by 1\% compared to the case in which they are neglected. 

\begin{figure}[!htb] 
  \centering
    \includegraphics[width=0.98\linewidth]{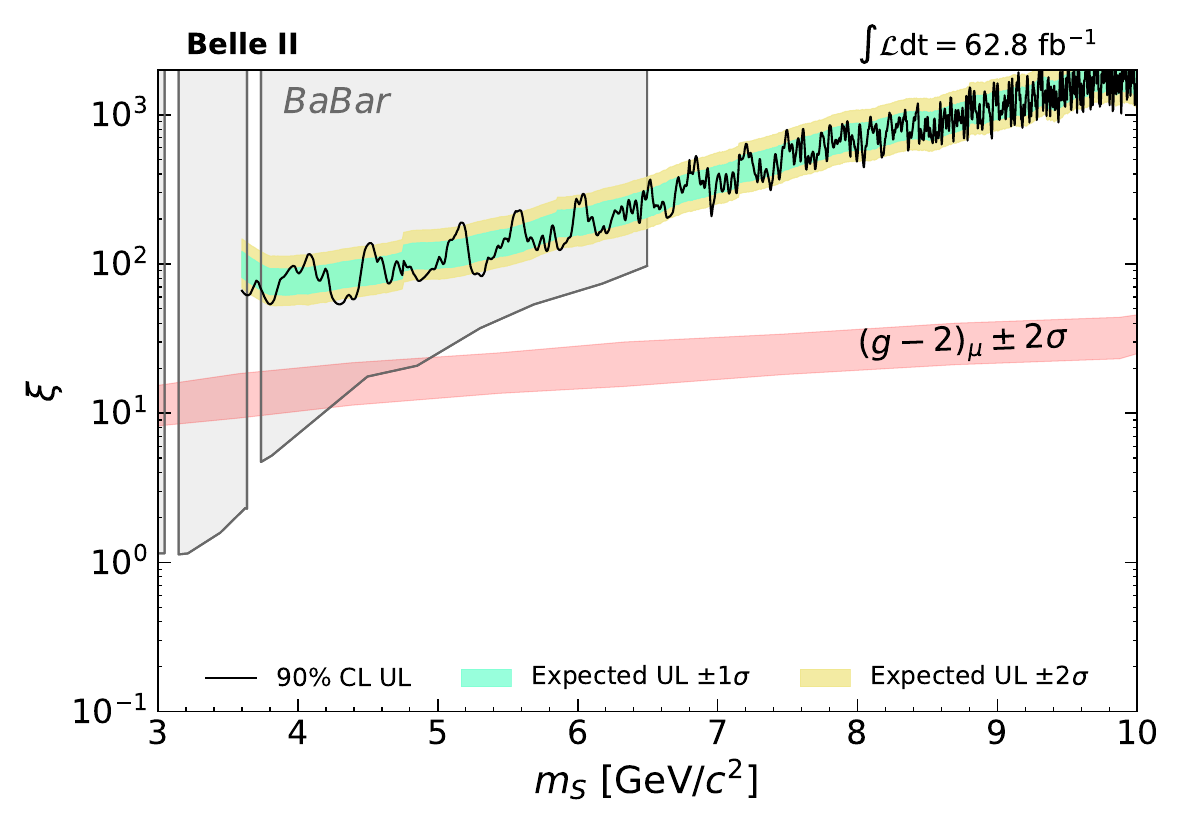}
    \includegraphics[width=0.98\linewidth]{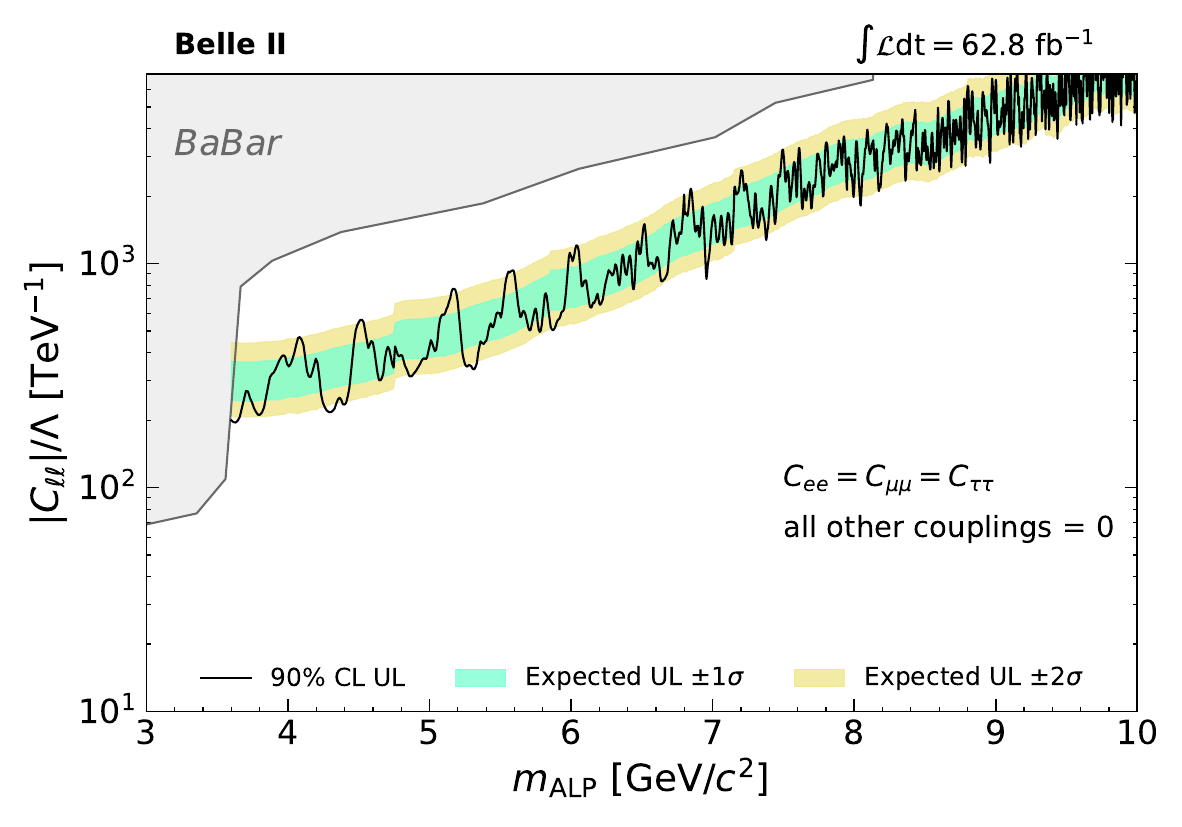}
    \caption{Observed 90\% CL upper limits and corresponding expected limits as functions of mass on ({\it top}) the leptophilic scalar coupling $\xi$, and on ({\it bottom}) the ALP coupling to leptons $|C_{\ell\ell}|/\Lambda$  
    in the hypothesis of equal couplings to the three lepton families and zero couplings to 
all other particles. 
Also shown are ({\it top}) constraints for $S$ decaying in electrons or muons from a \babar\ search~\cite{PhysRevLett.125.181801} and ({\it bottom}) constraints for an ALP decaying to leptons from a reinterpretation~\cite{Bauer_2017,bauer2022flavor} of \babar\ searches. The red band in the top plot shows the region that explains the muon anomalous magnetic moment $(g-2)_{\mu} \pm 2\sigma$.} 
    \label{fig:UL_couplings}
\end{figure}

The cross-section results are translated into upper limits on the coupling constant $g^\prime$ of the \lmultau\ model~\cite{supplemental_reference}, 
on the  coupling strength $\xi$ of the leptophilic  scalar $S$, and on the coupling $|C_{\ell\ell}|/\Lambda$
for an ALP decaying to leptons: values as low as 2.5$\times 10^{-2}$, 51, and 200 TeV$^{-1}$ are found, respectively. The last two are shown in Fig.~\ref{fig:UL_couplings} as functions of the resonance mass. 
For the leptophilic  scalar model, we constrain the coupling $\xi$ to be smaller than approximately 200
for masses above 6.5~\gevcc, which are  the first results in that region.
For the model with the ALP decaying to leptons, these are the first results for the ALP-$\tau$ coupling.

In summary, we search for a resonance decaying to \tautau\ in \mmtt\ events in a data sample of $e^{+}e^{-}$ collisions at 10.58~GeV collected by Belle~II in 2019--2020, corresponding to an integrated luminosity of 62.8~fb$^{-1}$. We find no significant excess above the background and set upper limits on the cross section, ranging from 0.7~fb to 24~fb, for masses between 3.6 and 10~\gevcc. We derive exclusion limits on the couplings for three different models: the \lmultau\ model; a leptophilic scalar model, for which we probe for the first time masses above 6.5~\gevcc; and a model with an ALP decaying to leptons, for which we set world-leading limits over the entire mass range considered. 

We thank Andrea Thamm for helpful conversations on the axion-like particle.

This work, based on data collected using the Belle II detector, which was built and commissioned prior to March 2019, was supported by
Science Committee of the Republic of Armenia Grant No.~20TTCG-1C010;
Australian Research Council and Research Grants
No.~DP200101792, 
No.~DP210101900, 
No.~DP210102831, 
No.~DE220100462, 
No.~LE210100098, 
and
No.~LE230100085; 
Austrian Federal Ministry of Education, Science and Research,
Austrian Science Fund
No.~P~31361-N36
and
No.~J4625-N,
and
Horizon 2020 ERC Starting Grant No.~947006 ``InterLeptons'';
Natural Sciences and Engineering Research Council of Canada, Compute Canada and CANARIE;
National Key R\&D Program of China under Contract No.~2022YFA1601903,
National Natural Science Foundation of China and Research Grants
No.~11575017,
No.~11761141009,
No.~11705209,
No.~11975076,
No.~12135005,
No.~12150004,
No.~12161141008,
and
No.~12175041,
and Shandong Provincial Natural Science Foundation Project~ZR2022JQ02;
the Czech Science Foundation Grant No.~22-18469S;
European Research Council, Seventh Framework PIEF-GA-2013-622527,
Horizon 2020 ERC-Advanced Grants No.~267104 and No.~884719,
Horizon 2020 ERC-Consolidator Grant No.~819127,
Horizon 2020 Marie Sklodowska-Curie Grant Agreement No.~700525 ``NIOBE''
and
No.~101026516,
and
Horizon 2020 Marie Sklodowska-Curie RISE project JENNIFER2 Grant Agreement No.~822070 (European grants);
L'Institut National de Physique Nucl\'{e}aire et de Physique des Particules (IN2P3) du CNRS
and
L'Agence Nationale de la Recherche (ANR) under grant ANR-21-CE31-0009 (France);
BMBF, DFG, HGF, MPG, and AvH Foundation (Germany);
Department of Atomic Energy under Project Identification No.~RTI 4002,
Department of Science and Technology,
and
UPES SEED funding programs
No.~UPES/R\&D-SEED-INFRA/17052023/01 and
No.~UPES/R\&D-SOE/20062022/06 (India);
Israel Science Foundation Grant No.~2476/17,
U.S.-Israel Binational Science Foundation Grant No.~2016113, and
Israel Ministry of Science Grant No.~3-16543;
Istituto Nazionale di Fisica Nucleare and the Research Grants BELLE2;
Japan Society for the Promotion of Science, Grant-in-Aid for Scientific Research Grants
No.~16H03968,
No.~16H03993,
No.~16H06492,
No.~16K05323,
No.~17H01133,
No.~17H05405,
No.~18K03621,
No.~18H03710,
No.~18H05226,
No.~19H00682, 
No.~22H00144,
No.~22K14056,
No.~23H05433,
No.~26220706,
and
No.~26400255,
the National Institute of Informatics, and Science Information NETwork 5 (SINET5), 
and
the Ministry of Education, Culture, Sports, Science, and Technology (MEXT) of Japan;  
National Research Foundation (NRF) of Korea Grants
No.~2016R1\-D1A1B\-02012900,
No.~2018R1\-A2B\-3003643,
No.~2018R1\-A6A1A\-06024970,
No.~2019R1\-I1A3A\-01058933,
No.~2021R1\-A6A1A\-03043957,
No.~2021R1\-F1A\-1060423,
No.~2021R1\-F1A\-1064008,
No.~2022R1\-A2C\-1003993,
and
No.~RS-2022-00197659,
Radiation Science Research Institute,
Foreign Large-Size Research Facility Application Supporting project,
the Global Science Experimental Data Hub Center of the Korea Institute of Science and Technology Information
and
KREONET/GLORIAD;
Universiti Malaya RU grant, Akademi Sains Malaysia, and Ministry of Education Malaysia;
Frontiers of Science Program Contracts
No.~FOINS-296,
No.~CB-221329,
No.~CB-236394,
No.~CB-254409,
and
No.~CB-180023, and SEP-CINVESTAV Research Grant No.~237 (Mexico);
the Polish Ministry of Science and Higher Education and the National Science Center;
the Ministry of Science and Higher Education of the Russian Federation,
Agreement No.~14.W03.31.0026, and
the HSE University Basic Research Program, Moscow;
University of Tabuk Research Grants
No.~S-0256-1438 and No.~S-0280-1439 (Saudi Arabia);
Slovenian Research Agency and Research Grants
No.~J1-9124
and
No.~P1-0135;
Agencia Estatal de Investigacion, Spain
Grant No.~RYC2020-029875-I
and
Generalitat Valenciana, Spain
Grant No.~CIDEGENT/2018/020;
National Science and Technology Council,
and
Ministry of Education (Taiwan);
Thailand Center of Excellence in Physics;
TUBITAK ULAKBIM (Turkey);
National Research Foundation of Ukraine, Project No.~2020.02/0257,
and
Ministry of Education and Science of Ukraine;
the U.S. National Science Foundation and Research Grants
No.~PHY-1913789 
and
No.~PHY-2111604, 
and the U.S. Department of Energy and Research Awards
No.~DE-AC06-76RLO1830, 
No.~DE-SC0007983, 
No.~DE-SC0009824, 
No.~DE-SC0009973, 
No.~DE-SC0010007, 
No.~DE-SC0010073, 
No.~DE-SC0010118, 
No.~DE-SC0010504, 
No.~DE-SC0011784, 
No.~DE-SC0012704, 
No.~DE-SC0019230, 
No.~DE-SC0021274, 
No.~DE-SC0022350, 
No.~DE-SC0023470; 
and
the Vietnam Academy of Science and Technology (VAST) under Grant No.~DL0000.05/21-23.

These acknowledgements are not to be interpreted as an endorsement of any statement made
by any of our institutes, funding agencies, governments, or their representatives.

We thank the SuperKEKB team for delivering high-luminosity collisions;
the KEK cryogenics group for the efficient operation of the detector solenoid magnet;
the KEK computer group and the NII for on-site computing support and SINET6 network support;
and the raw-data centers at BNL, DESY, GridKa, IN2P3, INFN, and the University of Victoria for off-site computing support.

\bibliography{files/references}

\end{document}


\title{\protect Search for a \texorpdfstring{$\tau^+\tau^-$}{tau} resonance in \texorpdfstring{\mmtt}{mmtt} events  with the Belle~II experiment}

\author{Belle II Collaboration}
\maketitle

\renewcommand{\thefigure}{S\arabic{figure}}
\setcounter{figure}{0}

\subsection*{Numerical results}

We provide a text file with numerical results of the observed cross section of \xtautau, where $X = Z',S, {\rm ALP}$, as well as of the observed 90\% CL upper limit on the cross section, $g'$, $\xi$, and $|C_{\ell\ell}|/\Lambda$ with $\ell = e, \mu, \tau$ as functions of the mass.

\subsection*{Discriminant variables}

Discriminating variables used as inputs of the MLP neural networks can be 
grouped in three classes: variables sensitive to the presence of a resonance in the final state; variables sensitive to the production mechanism, since the resonance is emitted as FSR off one of the two tagging muons; and variables sensitive to the presence of a $\tau^+\tau^-$ pair in the final state.

The first two classes contain variables expressed in the c.m. frame and mostly related 
to the kinematic properties of the two tagging muons.

Variables belonging to the third class are expressed in the reference frame where the recoil system against the two tagging muons is at rest (the \zprime\ rest frame, in case of signal).
 
\subsubsection*{Variables sensitive to the presence of a resonance in the final state.}
\begin{figure}[p] 
  \centering
    \includegraphics[width=0.7\linewidth]{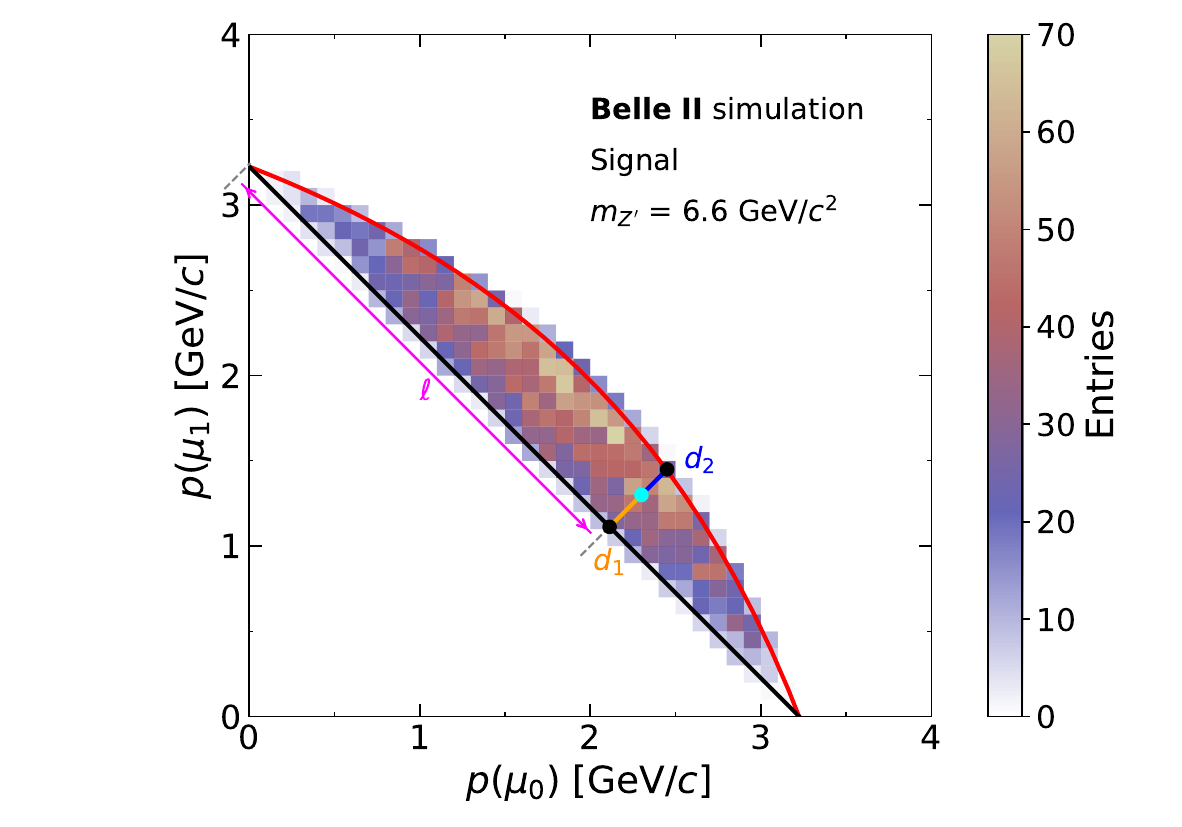}
    \includegraphics[width=0.7\linewidth]{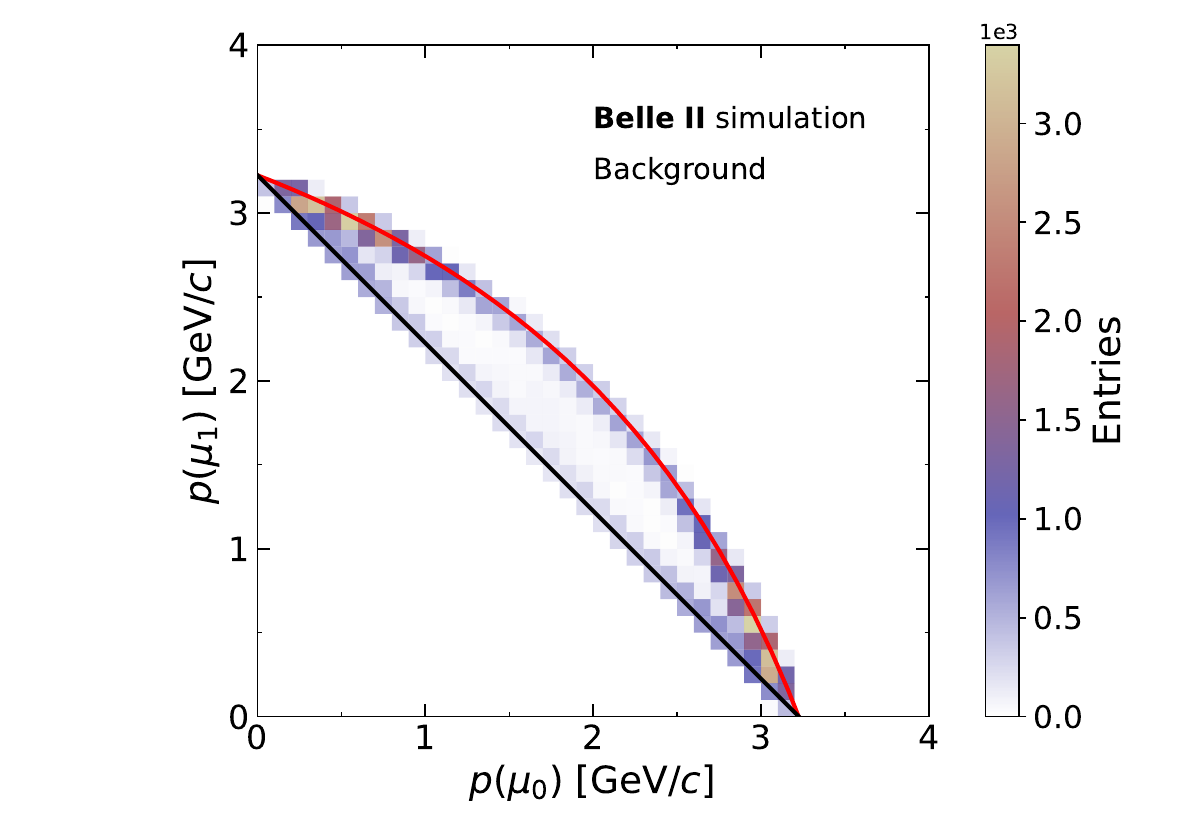}
    \caption{Distribution of the magnitudes of the momenta of the two tagging muons for simulated signal ({\it top}) and background ({\it bottom}) events. Signal is generated with a mass $m_{Z^\prime} = 6.6$ \gevcc. The reconstructed mass for signal and background is required to be in the interval 6.60 $\pm$ 0.05 \gevcc,
    that is within five times the mass resolution. The cyan point is a generic point of the distribution. The black points are the intersections of a straight line, perpendicular to the black line and passing through the cyan point, with the black line and the red hyperbola. The orange and blue segments are the distances $d_1$ and $d_2$, respectively. The magenta segment $l$ is the coordinate of the cyan point along the black line.} 
    \label{fig:p1vsp2}
\end{figure}

\begin{figure}[!hb] 
  \centering
    \includegraphics[width=0.8\linewidth]{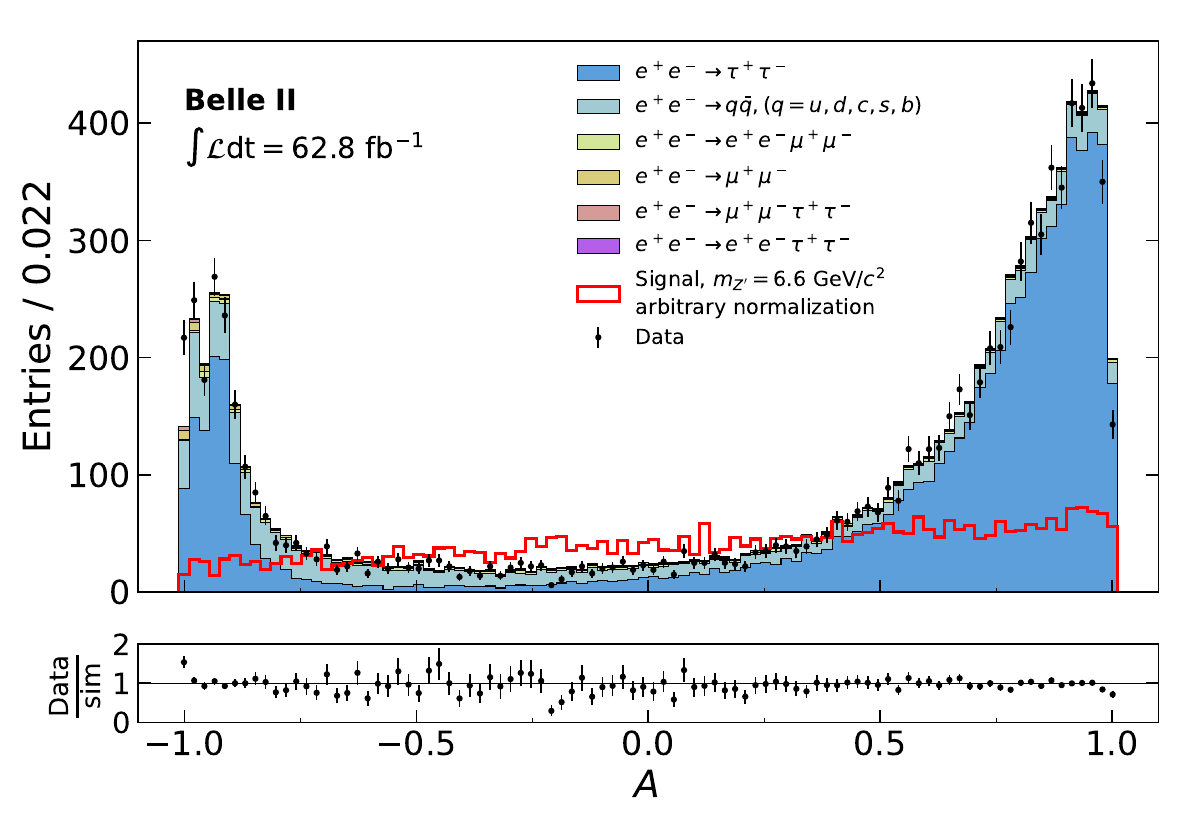}
    \caption{Distribution of the variable $A$ in the c.m. frame, for data and simulation. 
    $A$ is an asymmetry-like variable calculated from  the momenta of the two tagging muons. Contributions from the various simulated background processes are stacked. The simulation is normalized to the data luminosity, while the normalization of the signal is arbitrary.} 
    \label{fig:a}
\end{figure}

    \begin{itemize}
        \item The momenta of the two tagging muons are combined in two different variables, $A$ and $L$.
        The two-dimensional distribution of the magnitudes of the two tagging muons is shown in Fig.~\ref{fig:p1vsp2}, for signal and background: the distribution is confined within a straight line and a hyperbola, both analytically deducible from kinematic properties. Background processes populate the edges of the distribution, while signal is more uniformly distributed. To exploit this feature, an asymmetry-like discriminant variable $A$, shown in Fig.~\ref{fig:a}, is defined as $A=(d_1-d_2)/(d_1+d_2)$ where $d_1$ and $d_2$ are shown in Fig.~\ref{fig:p1vsp2}. 
        The second discriminant variable $L$ is related to the position of the generic point of the distribution along the straight line (see the segment $\ell$ in Fig.~\ref{fig:p1vsp2}), and is defined as $\ell$ scaled by the maximum recoil momentum kinematically reachable: background events cluster on the extremes of the distribution, while signal preferentially populates the central part. 
        \item The invariant mass of the two additional charged particles (other than the two tagging muons), and the sum of the magnitudes of their momenta.
    \end{itemize} 
 \subsubsection*{Variables sensitive to the production mechanism}
\begin{figure}[p] 
  \centering
    \includegraphics[width=0.8\linewidth]{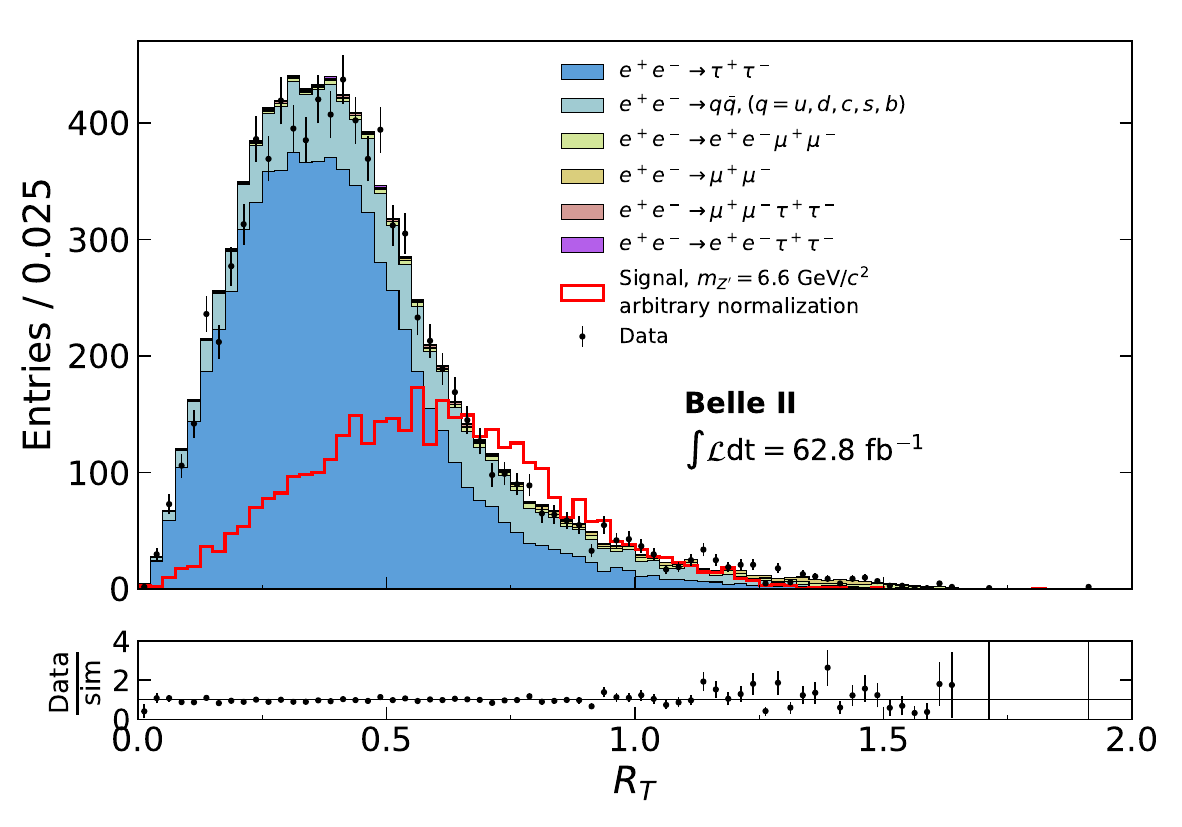}
    \caption{Distribution of the quadratic sum of the components of the recoil momentum transverse to the momentum direction of each of the two tagging muons in the c.m. frame,  
    divided by the maximum recoil momentum kinematically reachable, for data and simulation. 
    Contributions from the various simulated background processes are stacked. The simulation is normalized to the data luminosity, while the normalization of the signal is arbitrary.} 
    \label{fig:roverp}
\end{figure}

    The components of the recoil momentum
    transverse to the momentum direction of the tagging muon with maximum and minimum momentum, respectively, called  $p^{\mu-\mathrm{max}}_{T, \mathrm{recoil}}$ and $p^{\mu-\mathrm{min}}_{T, \mathrm{recoil}}$~\cite{PhysRevLett.124.141801}. In the case of signal, these are the transverse momenta of the $Z'$ with respect to the momentum direction of each of the two tagging muons. Since the $Z'$ is radiated off one of the two muons, these variables are sensitive to the signal FSR production. The $p^{\mu-\mathrm{max}}_{T, \mathrm{recoil}}$ and $p^{\mu-\mathrm{min}}_{T, \mathrm{recoil}}$ variables are combined to form two different variables that use 
    polar coordinates in the $p^{\mu-\mathrm{max}}_{T, \mathrm{recoil}}$ vs $p^{\mu-\mathrm{min}}_{T, \mathrm{recoil}}$ plane:  
    the quadratic sum scaled by the maximum recoil momentum kinematically reachable $R_T$, shown in Fig.~\ref{fig:roverp}, and the polar angle. 
    
\subsubsection*{Variables sensitive to the presence of a \texorpdfstring{$\tau^+\tau^-$}{tautau} pair in the final state.}
\begin{figure}[p] 
  \centering
    \includegraphics[width=0.8\linewidth]{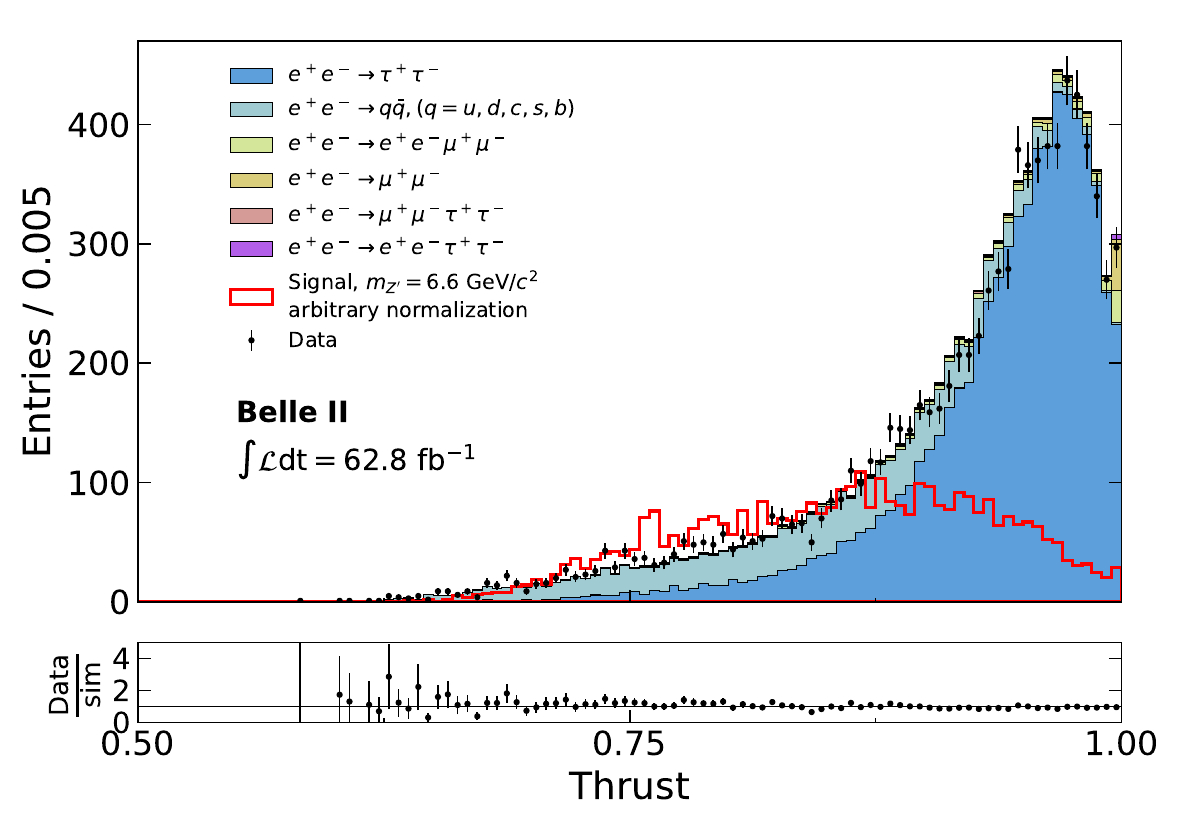}
    \caption{Thrust value distribution computed in the rest frame of the system recoiling against the two tagging muons, for data and simulation.
    Contributions from the various simulated background processes are stacked. The simulation is normalized to the data luminosity, while the normalization of the signal is arbitrary. } 
    \label{fig:thrust}
\end{figure}

\begin{figure}[p] 
  \centering
    \includegraphics[width=0.8\linewidth]{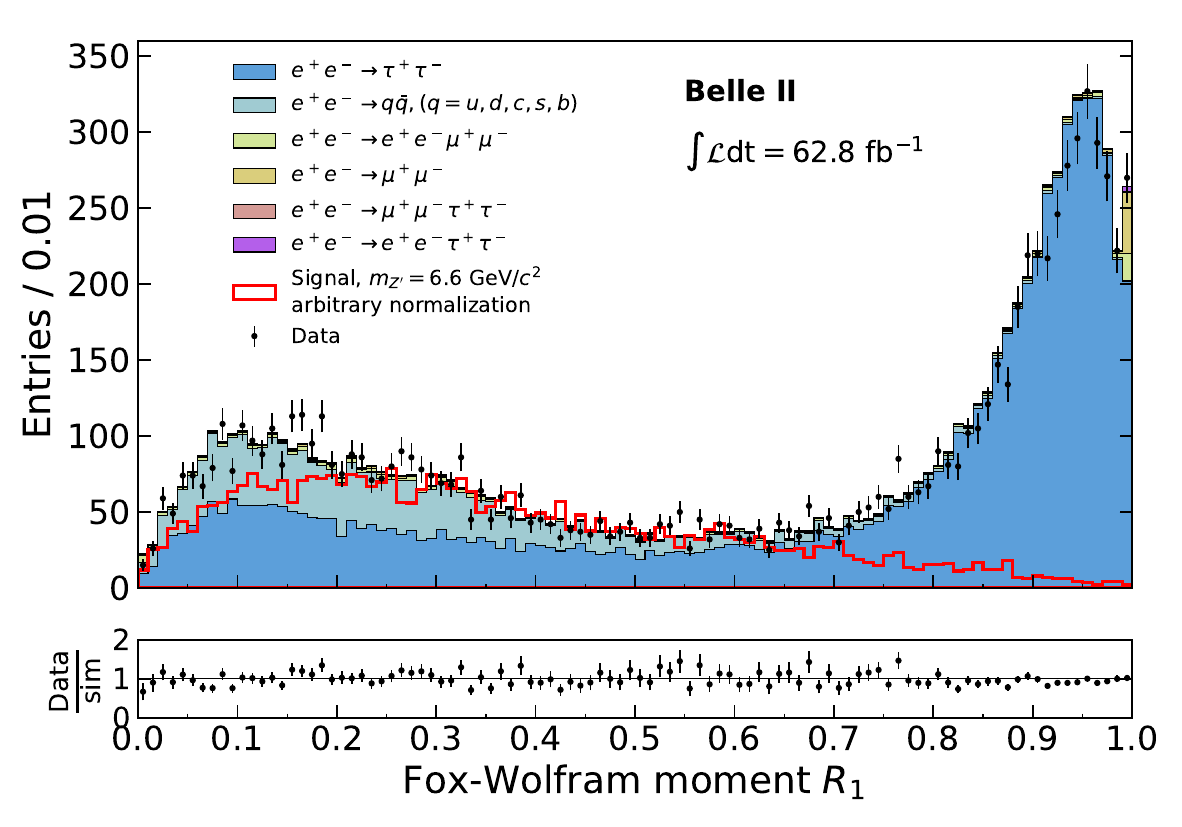}
    \caption{First Fox-Wolfram shape variable distribution computed in the rest frame of the system recoiling against the two tagging muons, for data and simulation. 
    Contributions from the various simulated background processes are stacked. The simulation is normalized to the data luminosity, while the normalization of the signal is arbitrary.} 
    \label{fig:foxr1}
\end{figure}

    \begin{itemize}
        \item Topological variables  such as the thrust value \cite{PhysRevLett.39.1587, Brandt:1964sa} (see Fig.~\ref{fig:thrust}), the first Fox-Wolfram moment shape variable~\cite{fox-wolfram} (see Fig.~\ref{fig:foxr1}), the angles between the thrust direction and the directions of each of the two tagging muons.
        \item Variables built exploiting the information of neutral pions, which are abundant in $\tau$ decays and expected both for the signal and for the 
        background. 
        The angles between each $\pi^0$ momentum and the direction of each of the two tagging muons are used to define two cases: the first case corresponds to topologies in which a $\pi^0$  is close or opposite to the directions of both tagging muons, the second case includes all the other topologies. For each 
        case, we consider the sum of the energies of all neutral pions as discriminant variable.
        Background events contribute mostly to the first case, while  
        signal is more uniformly distributed. 
        \item Variables built exploiting the rest-of-event (ROE), which is the system made of all charged and neutral particles except the two tagging muons: the difference between $M(4~\mathrm{tracks})$ and the sum of the ROE invariant mass, computed assuming the pion mass hypothesis for tracks and zero mass for neutrals, and the invariant mass of the two tagging muons; the difference between
        the total energy of the four-track system and the sum of the ROE energy and the total energy of the two tagging muons.
    \end{itemize}

\FloatBarrier
The following figures show the four most discriminating variables
before the MLP selection for data and simulation.
In all the cases, signal is generated with a mass $m_{Z^\prime} = 6.6$ \gevcc, and the reconstructed mass for signal and background is required to be in the interval 6.60 $\pm$ 0.05 \gevcc, that is within five times the mass resolution.
\vfill
\vfill

\FloatBarrier
\subsection*{Additional Figures}
\begin{figure}[!hp]  
  \centering
    \includegraphics[width=0.8\linewidth]{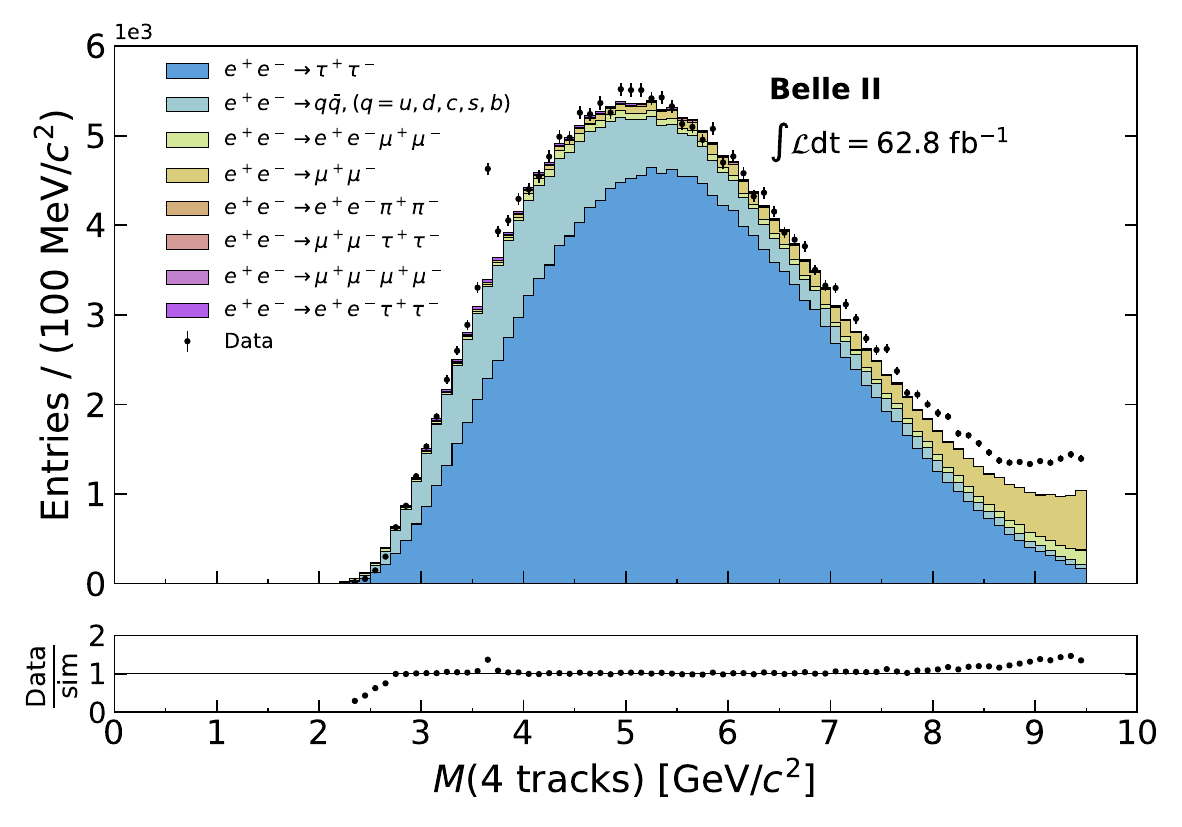}
    \caption{Observed four-track mass distribution, compared to the expectations of the simulation,  before the MLP selection and for $M(\mu\mu) >$~2~\gevcc. Contributions from the various simulated processes are stacked. Also visible in data is the $\psi(2S)$ resonance at about 3.7~GeV/$c^2$ through the decay chain $\psi(2S) \to J/\psi \, \,  \pi^+ \pi^-$ with $J/\psi \to \mu^+ \mu^-$, not present in the simulation. The large data-simulation discrepancy above 8~\gevcc\ is due to un-modelled ISR in simulation of four-lepton processes.} 
    \label{fig:m4tracks_premlp_cut}
\end{figure}

\begin{figure}[p] 
  \centering
    \includegraphics[width=0.8\linewidth]{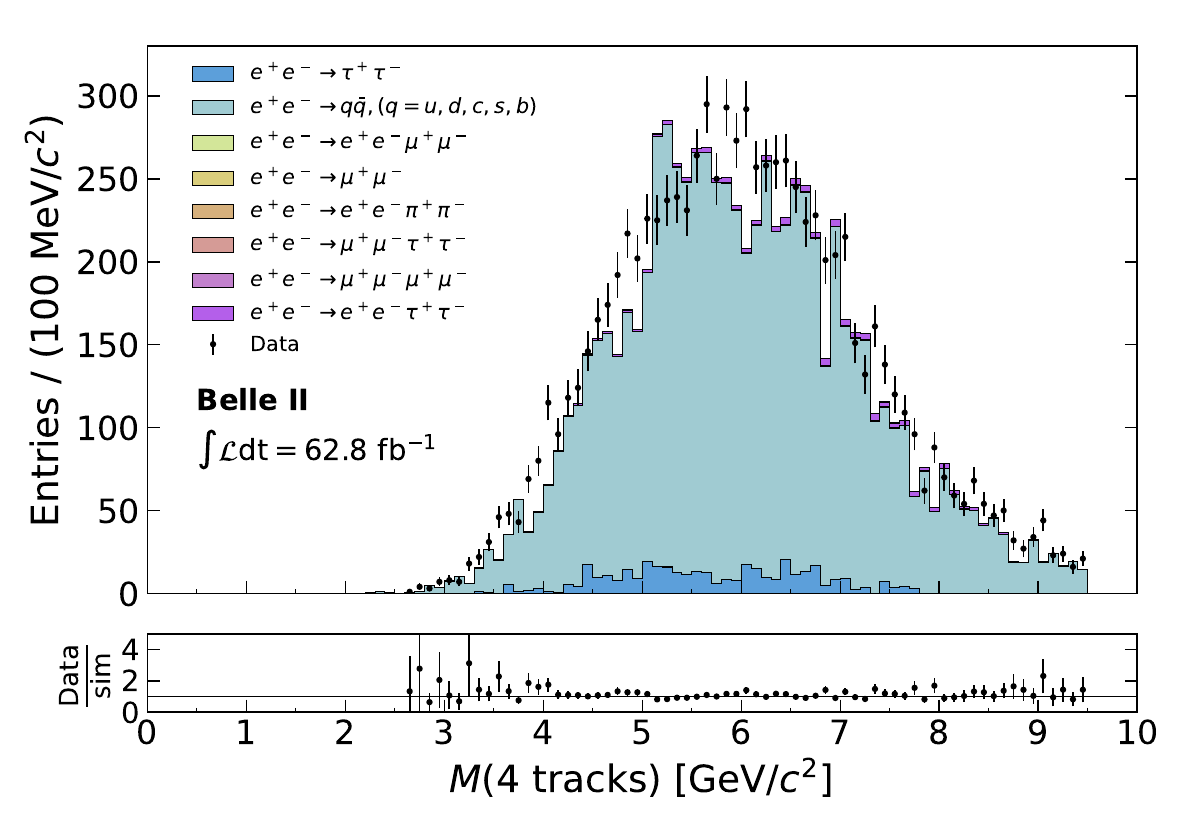}
    \caption{Observed four-track mass distribution, compared to the expectations of the simulation, in the pion-tagged control sample after all the selections and for $M(\pi\pi) >$~2~\gevcc. Contributions from the various simulated processes are stacked.} 
    \label{fig:4track_cs_postmlp_cut}
\end{figure}

Fig.~\ref{fig:m4tracks_premlp_cut} shows the distribution of \mtrk\ before the MLP selection, for $M(\mu\mu) >$~2~\gevcc. 
Fig.~\ref{fig:4track_cs_postmlp_cut} shows the distribution of \mtrk\ in the pion-tagged control sample after the MLP selection, for $M(\pi\pi) >$~2~\gevcc. These two distributions are used to understand the origin of the data-simulation discrepancies, as explained in the paper.

\begin{figure}[p]
\begin{center}
\includegraphics[width=0.75\linewidth]{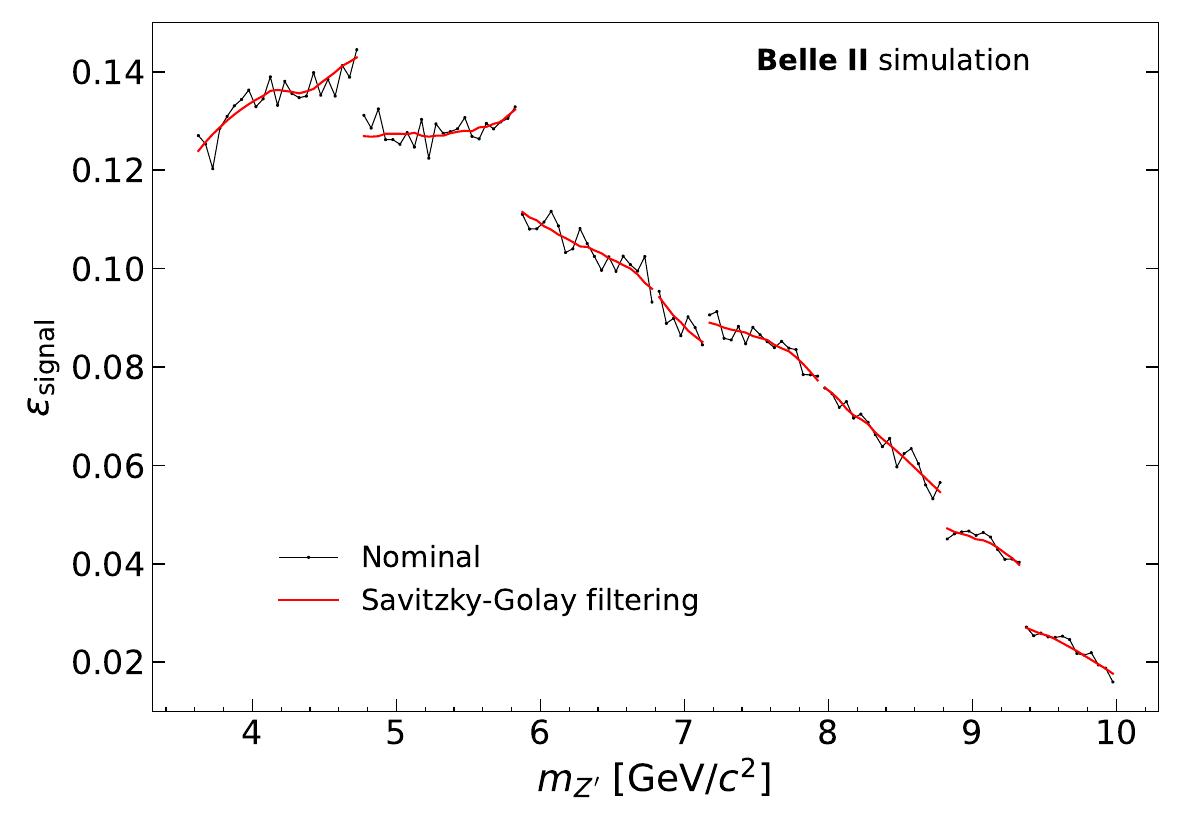}
\caption{\label{fig:eff} Signal efficiency as a function of the $Z'$ mass. In the analysis we use smoothed values obtained through the application of a Savitzky-Golay filter \cite{press1990savitzky} (red line).}
\end{center}
\end{figure}

Fig.~\ref{fig:eff} shows the signal efficiency as a function of the the $Z'$ mass after applying all the analysis selections.

\begin{figure}[p] 
  \centering
    \includegraphics[width=0.8\linewidth]{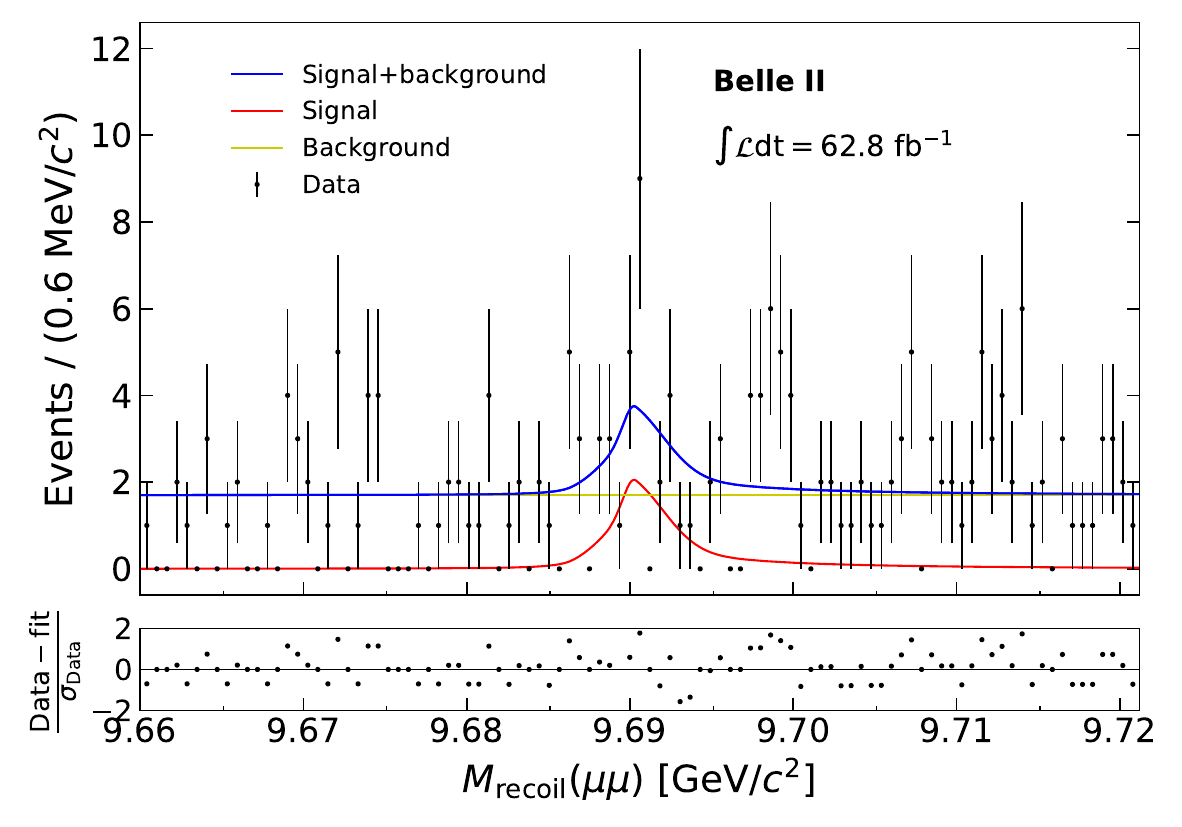}
    \caption{Fit for a \zprime\ mass hypothesis of 9.696 \gevcc, for which we obtain the maximum local significance $3.0 \,  \sigma$ (see text for definition). The bottom panel shows the difference between the observed and fitted events, divided by the statistical uncertainty of the former.} 
    \label{fig:best}
\end{figure}

Fig.~\ref{fig:best} shows the fit corresponding to the highest significance case.

\vfill
\vfill

\vfill

\FloatBarrier
\subsection*{Upper limits to the \texorpdfstring{\lmultau}{lmultau}\ model}

Upper limits on the coupling constants of the models are obtained from the upper limits on the cross sections, making use of the quadratic dependence. As an example, for the case of the \lmultau\ model, 

\begin{equation}
    {\rm UL}({g'})_{90\% \rm CL} = \sqrt{\frac{{g'}_{\rm ref}^{2}\cdot {\rm UL}(\sigma)_{90\% \rm CL}}{\sigma_{\rm ref}}},
\end{equation}
where $g^{\prime}_{\rm ref}$ is a  reference coupling constant used in the {\tt \textsc{MadGraph5@NLO}} generator to compute a reference cross section ($\sigma_{\rm ref}$).
\vfill
\begin{figure}[!htb] 
  \centering
    \includegraphics[width=0.8\linewidth]{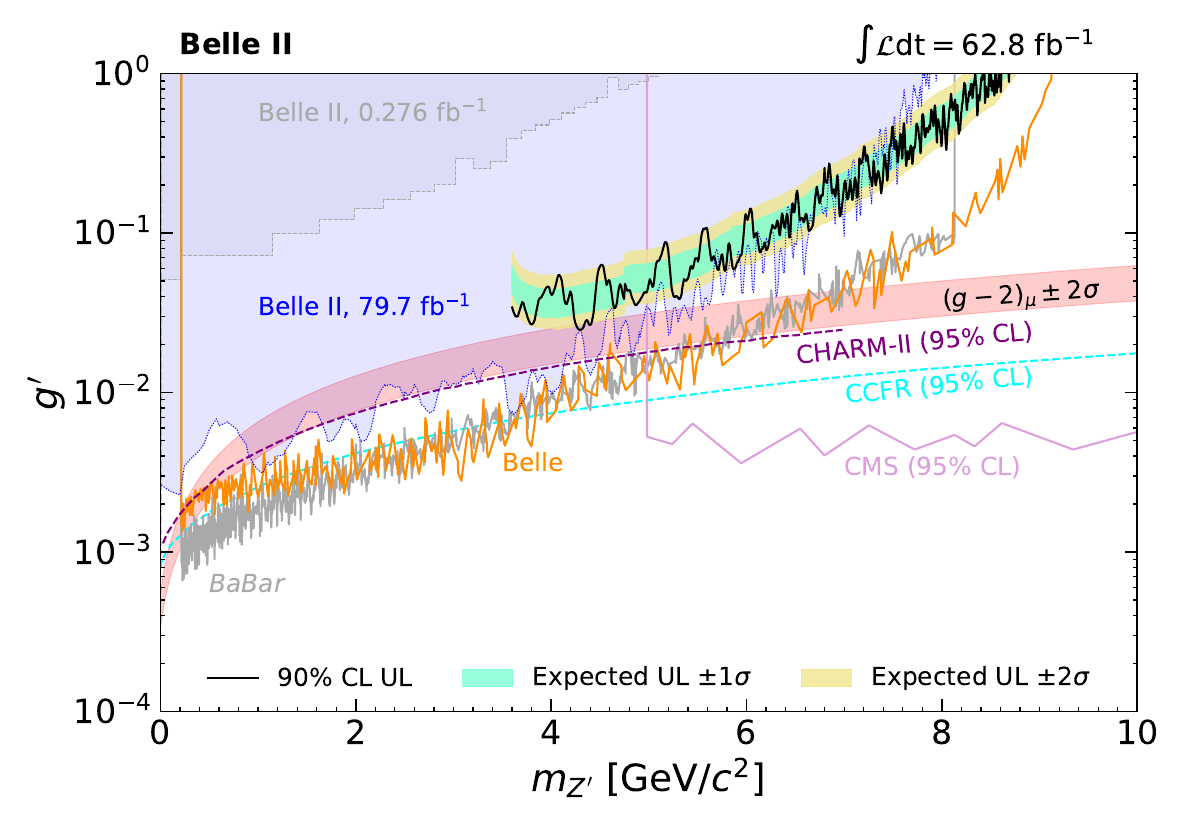}
    \caption{Observed 90\% CL upper limits on the $g^\prime$ coupling of the \lmultau\ model as a function of the \zprime\ mass. Also shown are constraints from  Belle~II~\cite{PhysRevLett.124.141801,PhysRevLett.130.231801}  for invisible \zprime\ decays, and from \babar~\cite{TheBABAR:2016rlg}, Belle~\cite{PhysRevD.106.012003}, and CMS~\cite{cms} (95\% CL) searches for \zprime\ decays to muons, along with constraints (95\% CL) derived from the trident production in neutrino experiments~\cite{PhysRevLett.113.091801,PhysRevLett.107.141302,kamada2018self}. The red band shows the region that could explain the observed value (within two standard deviations) of the muon anomalous magnetic moment~\cite{PhysRevLett.126.141801}.} 
    \label{fig:gp_lmultau}
\end{figure}
\vfill

\newpage
\bibliography{files/references}